\documentclass[letterpaper,12pt]{article}
\usepackage{epsfig}
\usepackage{graphicx}
\usepackage{amsmath}
\usepackage{amsfonts}
\usepackage{dsfont}
\usepackage{amssymb}
\usepackage{mathrsfs}
\usepackage{subfigure}
\usepackage{wasysym}
\newcommand{\be}{\begin{equation}}
\newcommand{\ee}{\end{equation}}
\newcommand{\bea}{\begin{eqnarray}}
\newcommand{\eea}{\end{eqnarray}}

\title{\textbf{Modified Pati--Salam Model from $Z_7$ orbifolded AdS/CFT}}
\author{James B.~Dent$^{1,}$\thanks{Electronic address: james.b.dent@vanderbilt.edu} ,
        Pasquale L.~Iafelice$^{2,}$\thanks{Electronic address: iafelice@bo.infn.it} 
        and
        Thomas W.~Kephart$^{1,}$\thanks{Electronic address: tom.kephart@gmail.com}\\[6pt]  
      \textit{\normalsize{$^1$Department of Physics and Astronomy, Vanderbilt University,}}\\[0pt] 
                \textit{\normalsize{ Nashville, TN 37235, USA}}\\[3pt]
      \textit{\normalsize{$^2$Dip.~di Fisica - Universit\`a di
                       Bologna and INFN - Sezione di Bologna,}}\\[0pt]
         \textit{\normalsize{via Irnerio 46, 40126 Bologna, Italy}}
          }

\date{\today}
\begin{document}
\maketitle
\begin{abstract}
\noindent
We consider models built on $AdS_5\otimes S^5/\Gamma$ orbifold compactifications of the type $IIB$ superstring, where $\Gamma$ is the abelian group $Z_n$. An attractive three family $\mathcal{N}=0$ SUSY model is found for $n=7$  that is a modified Pati--Salam Model which unifies at about 5 TeV and reduced to the Standard Model after symmetry breaking.
\end{abstract}
%
\section{Introduction}
\label{intro}
\setcounter{equation}{0}
The simplest compactification of a ten-dimensional superstring on a product of an AdS space with a five-dimensional spherical manifold leads to an $\mathcal{N} = 4$ $SU(N)$ supersymmetric gauge theory, well known to be conformally invariant \cite{Mandelstam:1982cb}. By replacing the manifold
$S^5$ by an orbifold $S^5/\Gamma$ one arrives at a theory with less supersymmetries corresponding to $\mathcal{N} = 2$, $1$ or $0$
 depending \cite{Kachru:1998ys} on whether: (i) $\Gamma \subset  SU(2)$, (ii) $\Gamma \subset  SU(3)$ but not in $SU(2)$, or (iii) $\Gamma \subset  SU(4)$ but not in $SU(3)$ respectively, where $\Gamma$ is in all cases a subgroup of $SU(4) \sim SO(6)$, the isometry of the $S^5$ manifold (for a review see \cite{Frampton:2007fr}). It was conjectured in \cite{Maldacena:1997re} that such $SU(N)$ gauge theories are conformal in the $N \rightarrow \infty$ limit. In \cite{Frampton:1999wz} it was conjectured that at least a subset of the resultant nonsupersymmetric $\mathcal{N} = 0$ theories are conformal even for finite $N$ and that one of this subset may provide the correct extension of the Standard Model.

Recently,  all $\mathcal{N}=0$ and $\mathcal{N}=1$ SUSY models have been classified \cite{Kephart:2001qu, Kephart:2004qp} that come from orbifolding $AdS_5\otimes S^5 $ with an abelian group $\Gamma$ of order less than $12$, where $\Gamma$ embeds irreducibly in the $SU(4)$ isometry or in an $SU(3)$ subgroup of the $SU(4)$ isometry, respectively. This means that, to achieve $\mathcal{N}=0$, $rep(\Gamma) \rightarrow {\bf 4}$ of $SU(4)$ must be embedded as ${\bf 4} = ({\bf r})$, where ${\bf r}$ is a nontrivial four dimensional representation of $\Gamma$; for $\mathcal{N} = 1$, $rep(\Gamma) \rightarrow {\bf 4}$ of $SU(4)$ must be embedded as ${\bf 4} = ({\bf 1, r})$, where ${\bf 1}$ is the trivial singlet of $\Gamma$ and ${\bf r}$ is nontrivial.

We want to focus on non-supersymmetric Pati-Salam (PS) type models (for a SUSY version see \cite{Dent:2007eu}). One motivation for studying the nonSUSY case is that
the need for supersymmetry is less clear in CFT as: (1) the hierarchy problem is absent or ameliorated, (2) the difficulties involved in breaking the remaining $\mathcal{N} = 1$ SUSY can be avoided if the orbifolding already results in $\mathcal{N} = 0$ SUSY, and (3) many of the positive effects of SUSY are still present in the theory, although just hidden. 

For $\mathcal{N}=0$ the fermions are given by $\sum_i {\bf 4} \otimes R_i$ and the scalars by $\sum_i {\bf 6} \otimes R_i$ where the set ${R_i}$ runs over all the irreps of $\Gamma$. For $\Gamma$ abelian, for example $\Gamma=Z_n$, the irreps are all one dimensional and as a consequence of the choice of $N$ in the 1/N expansion, the gauge group  is $SU^n(N)$ \cite{Lawrence:1998ja}.

In this paper, starting from the classification of Kephart and Pas (2004), we have searched for a minimal (respect to the order of $\Gamma=Z_n$) nonSUSY model that have SM particles as a subset of its particle content. To do this we have  used symmetry breaking paths that contain the Pati-Salam (PS) group as a subgroup before reaching the SM. The minimal model of this type has symmetry group $SU^7(4)$, hence orbifolding group is $Z_7$, as we will discuss.

The running of the coupling constants predicted by the model depends strongly on the scalar content. In fact, since there are scalars in addition to the usual SM Higgs sector, they can contribute to the running of the beta functions. After a presentation of the model and of the SSB chain that leads to the SM particle content, we show that, with the use of a judicious choice of the scalar sector, unification can be achieved at the scale $M_{GUT}\sim 10^3$ GeV.  We then conclude with a few comments on the phenomenology of the model including proton decay constraints and dark matter.

\section{Description of the model}
\setcounter{equation}{0}
We have systematically gone through all chiral models with $\Gamma=Z_n$. All fail to have a PS type intermediate stage until $n=7.$  Hence
after considerable exploration, we are lead to choose  $\Gamma=Z_7$ and $N=4$ with orbifold group embedding ${\bf 4}=(\alpha,\alpha,\alpha^2,\alpha^3)$. This yield an $\mathcal{N}=0$ SUSY model based on the gauge group $SU(4)^7$. The particle spectrum of the unbroken theory at the string scale is given by the fermion states
\begin{eqnarray}
 2\left[(4\bar{4}11111)_F  + \cdots \; \right]
 +[(41\bar{4}1111)_F +   \cdots]
 +[(411\bar{4}111)_F    + \cdots ]
  \nonumber
\end{eqnarray}
and scalars
\begin{eqnarray}
 2[(41\bar{4}1111)_S +   \cdots]
 +4\left[(411\bar{4}111)_S +  \cdots \;\right]
 +4\left[(4111\bar{4}11)_S +   \cdots \;\right]
 +2[(41111\bar{4}1)_S + \cdots]
  \nonumber
\end{eqnarray}
of $SU(4)^7$, where the dots mean cyclic permutations. $SU(4)^7$ is broken down to $SU(4)^3$ via diagonal subgroups by sequentially assigning vacuum expectation values (VEVs) to $(141\bar{4}111)_S$, $(114\bar{4}11)_S$, $(114\bar{4}1)_S$ and $(114\bar{4})_S$, which leaves chiral fermions in the following bifundamental representations
\begin{equation}
 3\left[(4\bar{4}1)+(14\bar{4})+(\bar{4}14)\right]_F
 \label{SU(4)^3fermion}
\end{equation}
and scalars  
\begin{eqnarray}
 4\left[(4\bar{4}1)+h.c]_S\;
 +\; 8[(41\bar{4})+h.c.\right]_S\;
 +\; 16\left[(14\bar{4})+h.c.\right]_S\,\nonumber\\
 21\left[(1,1,15)\right]_S \;
 +\; 3(15,1,1)_S\;\,\\
 28\left[(111)\right]_S\nonumber.
\end{eqnarray}
We continue the chain of spontaneous symmetry breaking toward the Pati--Salam model with a VEV for the $(4\bar{4}1)$ of the form
$$\left(\!\!
\begin{array}{llll}1 & 0 & 0 &  \,\, 0 \\ 0 & 1 & 0 & \,\, 0 \\ 0 & 0 & 1 & \,\, 0 \\ 0 & 0 & 0 & -3
\end{array}
\!\!\right)\!\!.$$
 This breaks the symmetry to $SU(3)\otimes SU(3)\otimes SU(4)\otimes U(1)_A$ (see \cite{Kephart:2001ix, Kephart:2006zd} for a detailed study of the phenomenology of this model without $U(1)_A$ charge) and gives three  $U(1)_A$ neutral $(3\bar{3}1)_0$ scalars. Finally, giving a VEV of the form
 $$\left(\!\!
 \begin{array}{lll}1 &  0 &  \,\, 0 \\ 0 &  1 & \,\, 0 \\ 0 &  0 & -2
 \end{array}
\!\! \right)$$
to one of these $(3\bar{3}1)_0$s, we arrive at the gauge group $SU(2)_L\otimes SU(2)_R\otimes SU(4)_C\otimes U(1)_A\otimes U(1)_B$ that resemble the Pati--Salam model group $SU(2)_L \otimes SU(2)_R\otimes SU(4)_C$. A this stage the scalar content is given by Table~\ref{PSscalar}.
\begin{table}[h!]
\centering
 \begin{tabular}{|c|c|}
 \hline 
 \multicolumn{2}{|c|}
 {{\bf Scalars} of $SU(2)_L\otimes SU(2)_R\otimes SU(4)_C\otimes U(1)_A\otimes U(1)_B$}\\
 \hline  \hline
 $ $ & $ $\\
  $24\left[(11\bar{4})_{1,0}+ (11\bar{4})_{-1/3,1} + h.c.\right]$ & $8\left[ (21\bar{4})_{-1/3,-1/2}+h.c.\right]$  \\
 $ $ & $ $\\
  $16\left[(12\bar{4})_{-1/3,-1/2}+h.c.\right]$  &   $21\left[(1,1,15)_{0,0}\right]$  \\ 
 $ $ & $ $\\
  $8(221)_{0,0} + 3\left[(211)_{-4/3,-1/2} + h.c.\right]$   &  $3\left[(211)_{0,-3/2}+ h.c.\right]$\\
 $ $ & $ $\\
  $6\left[(121)_{4/3,1/2} + h.c.\right]$    &   $6\left[(121)_{0,-3/2} + h.c.\right]$\\
 $ $ & $ $\\
  $9\left[(111)_{4/3,-1}+h.c.\right]$ & $48\left[(111)_{0,0} \right]$ + $3(131)_{0,0}$\\
 $ $ & $ $\\
\hline
 \end{tabular}
 \caption{Scalars of the generalized Pati-Salam model.}
 \label{PSscalar}
\end{table}

In order to arrive at the Standard Model we will break $SU(4)_C\rightarrow SU(3)_C\times U(1)_X$ and $SU(2)_R\rightarrow U(1)_Z$.  This can be accomplished by giving a VEV to a scalar in the $(12\bar{4})_{-1/3,-1/2}$ representation, which leads to the group $SU(2)_L\times SU(3)_C$ and three $U(1)$ factors. More precisely this would result in four $U(1)$ factors, but one linear combination is broken due to the non-zero $U(1)$ charges of $(12\bar{4})_{-1/3,-1/2}$. Nevertheless we write these four charge as superscripts in order to fix the normalization later. \\

Under the group structure $SU(2)_L \times SU(3)_C\times U(1)_X\times U(1)_Z\times U(1)_A\times U(1)_B$, the scalar state $(12\bar{4})_{-1/3,-1/2}$ decomposes to $(1\bar{3})_{1/3,1,-1/3,-1/2} + (11)_{-1,1,-1/3,-1/2} + (1\bar{3})_{1/3,-1,-1/3,-1/2} + (11)_{-1,-1,-1/3,-1/2}$, while the scalar state $(11\bar{4})_{-1/3,1}$ decomposes to $(1\bar{3})_{1/3,0,-1/3,1} + (11)_{-1,0,-1/3,1}$.  Therefore giving a VEV to $(11)_{-1,1,-1/3,-1/2}$, $(11)_{-1,0,-1/3,1}$ and the additional scalar $(11)_{0,0,4/3,-1}$, can break $SU(2)_L\times SU(2)_R \times SU(4)_C \times  U(1)_A\times U(1)_B$ down to $SU(2)_L \times SU(3)_C $, along with a single $U(1)$ formed by a linear combination of four $U(1)$ factors.  

Since we are breaking three combinations of four $U(1)$ charges we must ensure that there exists a normalization pattern that will result in the remaining $U(1)$ being the usual hypercharge of the Standard Model.  Starting from the well know Gell Mann-Nishima relation $Q=T_3+Y$, with $Q$ being the electric charge, $T_3$ the third isospin component and $Y$ the hypercharge, we can choose a suitable normalization of the charges $A$, $B$, $X$, and $Z$ of the form
\begin{equation}
 xX+zZ+aA+bB=Y,\qquad x=\frac{1}{4},\; z=\frac{1}{2}, \;a=\frac{1}{4}, \; b=\frac{1}{3}\,.
 \label{ChargeNorm}
\end{equation}
This completes the chain of symmetry breaking reaching the Standard Model gauge group $U(1)_Y\otimes SU(2)_L \otimes SU(3)_C$.  The fermion content from Eq.~\eqref{SU(4)^3fermion} becomes three chiral families of the Standard Model plus the following vectorlike states: eight adjoints of $SU(3)_C$ and one adjoint of $SU(2)_L$. Moreover there are numerous right handed neutrinos.  The scalar content is given in Table~\ref{SMscalar}.
\begin{table}[h!]
\centering
 \begin{tabular}{|c|c|}
 \hline 
 \multicolumn{2}{|c|}
 {{\bf Scalars} of $U(1)_Y\otimes SU(2)_L \otimes SU(3)_C$}\\
 \hline  \hline
 $ $ & $ $\\
  $8\left[(23)_{1/6}+h.c\right]$ &   $84\left[(1\bar{3})_{1/3}+ h.c.\right]$ \\
 $ $ & $ $\\
  $16\left[(13)_{2/3}+h.c.\right]$  &   $22\left[(21)_{1/2}+h.c.\right]$  \\ 
 $ $ & $ $\\
  $31\left[(11)_{-1}+h.c.\right]$   & $21\left[(18)_{0}\right]$ \\
 $ $ & $ $\\
  $237(11)_0$   &   \\
 $ $ & $ $\\
\hline
 \end{tabular}
  \caption{Scalars at the Standard Model level.}
  \label{SMscalar}
 \end{table}
 
\hspace{2cm}

%
%
\section{Phenomenology}
In the previous section, the symmetry breaking of the initial $SU(4)^7$ towards to $SU(4)_R\otimes SU(4)_L\otimes SU(4)_C$ gauge group was performed by allowing the states
 $(141\bar{4}111)_S$, $(114\bar{4}11)_S$,  $(114\bar{4}1)_S$,   $(114\bar{4})_S$ to obtain VEVs.  This makes clear that $SU(4)_R$, $SU(4)_L$ and $SU(4)_C$ are embedded in diagonal subgroups $SU(4)^q$, $SU(4)^p$ and $SU(4)^r$ of $SU(4)^7$, respectively. We then embed all of $SU(2)_L$ in $SU(4)_L$, but for $U(1)_Y$ the embedding is slightly more complicated. We need to go back to Eq.(\ref{ChargeNorm}) to read the fraction of $U(1)_Y$ embedded in each of the $U(1)_{X,Z,A,B}$ factors. Considering also that we embed all of $U(1)_X$ in $SU(4)_C$, all of $U(1)_Z$ in $SU(4)_R$, $1/2$ of $U(1)_{A,B}$ in $SU(4)_L$ and the other $1/2$ in $SU(4)_R$, the ratio $\alpha_2/\alpha_1$ of the coupling constants turns out to be
$$
\frac{5}{3} \frac{\alpha_2}{\alpha_1}= \frac{\alpha_2}{\alpha_Y}= \frac{\frac{1}{4}r+\frac{1}{2}q+\frac{1}{4}\left(\frac{p+q}{2}\right)+\frac{1}{3}\left(\frac{p+q}{2}\right)}{p}$$ 
and $\sin^2\theta_W$ satisfies (see \cite{Frampton:2001xh} and references therein)
\begin{equation}
\sin^2\theta_W(M_{GUT})=
\frac{3}{3+5(\frac{\alpha_2}{\alpha_1})}=
\frac{24 p}
{6 r+ 31 p+ 19 q}\,.
\end{equation}
In our  $n=7$ model, $r=4$, $p=1$ and $q=2$ gives 
$$\sin^2\theta_W(M_{GUT})=8/31\simeq 0.26$$ 
and the unification scale $M_{GUT}$ is such that 
\begin{equation}
\frac{\alpha_3(M_{GUT})}{\alpha_2(M_{GUT})}=\frac{r}{p}=4
\end{equation}
together with 
\begin{equation}
\frac{\alpha_2(M_{GUT})}{\alpha_1(M_{GUT})}=\frac{3}{5}\frac{6r+7p+19q}{24p}=\frac{69}{40}.
\end{equation}
To find this energy scale we consider the renormalization-group evolution of the gauge couplings in leading order as given by
\begin{equation}
\label{RGE}
\alpha_i(Q)=\frac{1}{\alpha_i(Q')^{-1}+\frac{b_i}{2\pi}\ln\left(\frac{Q}{Q'}\right)},
\end{equation}
where $b_i$ are the one-loop contributions to the beta function coefficients
that are given in general by \cite{Machacek:1983tz}
\begin{equation}
b_i=\frac{11}{3}C_2(G)-\frac{4\kappa}{3}S_2(F)-\frac{1}{6}S_2(S)
\end{equation}
Here $n_F$ is the number of chiral families, $C_2(G)$ is the quadratic Casimir invariant for the gauge group $G$ and  $S_2(F)$ and $S_2(S)$ are the Dynkin indices for the fermion and scalar representations $F$ and $S$ respectively, and $\kappa$ is $\frac{1}{2}$ for Weyl fermions and 1 for Dirac fermions, see also \cite{Cheng:1973nv,Slansky:1981yr,Frampton:2001xh}. 
For the case at hand
\begin{eqnarray}
b_3&=&11-\frac{4}{3}n_F-\frac{1}{6}N_{ST},\\
b_2&=&\frac{22}{3}-\frac{4}{3}n_F-\frac{1}{6}N_{SD},\\
b_1&=&-\frac{4}{3}n_F-\frac{1}{10}\sum_{i=1}^n d_i q_i^2.
\end{eqnarray}
In $b_3$, $N_{ST}$ is the number of real scalar triplets, in $b_2$, $N_{SD}$ is the number of real scalar doublets, and in $b_1$ the sum runs over the scalar representation with  $U(1)$ charges $q_i$ of dimensions $d_i$. In our model $n_F=3$.

\begin{figure}[h!]
   \begin{center}
     \includegraphics[scale=1.3]{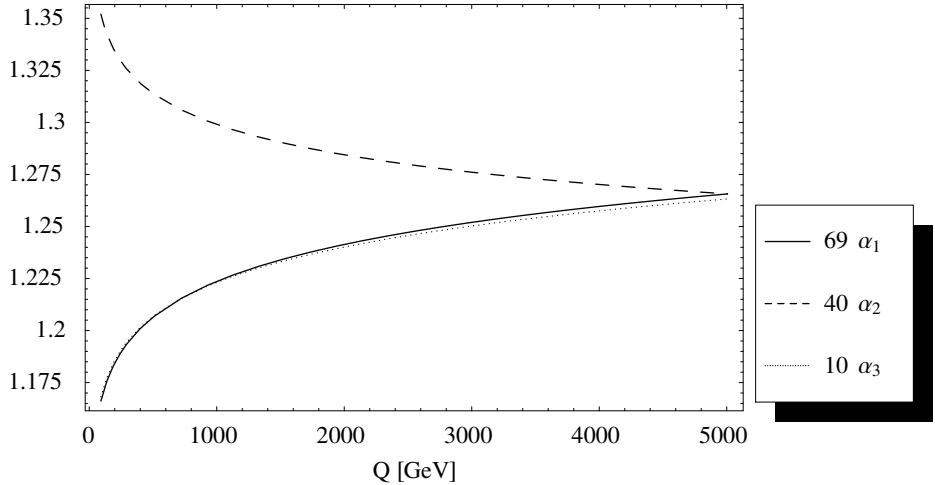}
   \end{center}
   \caption{Gauge coupling unification in the Modified Pati-Salam model. The curves has been rescaled as $69\alpha_1(Q)$, $40\alpha_2(Q)$ and $10\alpha_3(Q)$ in such a way that their ratio match to one at the unification scale. The plot is for values of $Q$ from $M_Z$ to $M_{GUT}$. Note that $SU(3)_C$ is no longer asymptotically free above the scalar triplet threshold, but it is asymptotically free at low energy as required.}
   \label{fig:CCrunning}
\end{figure}

 The experimental input values of the gauge couplings are \cite{Hagiwara:2002fs} 
\begin{equation}
\alpha_1(M_Z)=0.01014,\quad  \alpha_2(M_Z)=0.0338, \quad \alpha_3(M_Z)=0.118\,.
\end{equation}
We can choose the number of light scalar representations, i.e, use the $S_2$'s in the equations \eqref{RGE} as parameters to match ratio between the coupling constant at the GUT scale. As an example, this procedure leads to an unification scale $M_{GUT}=5.0 \times 10^{3}$ GeV,
for the choice of a single Higgs doublet plus 24 complex color triplet scalars of hypercharge 1/3. The evolution of the couplings from the weak to the unification scale is shown in Fig.~\ref{fig:CCrunning}. 
Changing the choice of light scalars adjusts the unification scale, but given the experimental input at low energy and the requirement of unification at a higher scale, we necessarily need many scalars to be light below the unification scale. 
Increasing the triplet scalar masses (they would probably already have been detected, at least indirectly, if they were at the weak scale) to a few hundred GeV would likewise increase the unification scale to the 6 TeV range. Using extra vectorlike fermions instead of scalars can achieve similar results and with fewer particles, since fermions contribute more strongly to the $\beta$ functions.

\section{Conclusions}
We have shown that it is possible to find a non-supersymmetric, ``minimal'', Pati-Salam type model based on the $AdS/CFT$  orbifold compactifications of type $IIB$ string theory on $AdS_5\otimes S^5/Z_7$. At the unification scale, this model contains bifundamental fermion and scalar representations of the gauge group $SU(4)^7$, where the one loop, and perhaps higher loop $\beta$ functions vanish, and conformality is partially, or fully restored.  The type of fields arising in such a model are constrained by the orbifold group yet we have shown that there exists the proper scalar content to allow spontaneous symmetry breaking to the Standard Model, as well as provide the usual Higgs sector of the Standard Model.   To achieve low scale unification, we require scalar content beyond what is found in the Standard Model Higgs sector. Conversely, the existence of such particles may be an indicator of low scale unification. (Similar results hold for extra vectorlike fermions.) The m!
 odel contains three families of chiral fermions with standard model charge assignments, but with no other chiral fermions at low energy. There are a sufficient number of right handed neutral singlet fermions at intermediate or higher mass to provide neutrino see saw masses. Proton decay is avoided as the model unifies into a modified Pati-Salam model at the intermediate scale $ M_{GUT}$ = 5 TeV.  Generically, the unification is lowered by keeping more
scalars light (similar results would hold if we replaced them with vectorlike fermions).  Since our model is not supersymmetric, there is no natural LSP dark matter candidate, but one can still expect other options to be available, e.g., axionic dark matter, although we will not explore these possibility here. 

\subsubsection*{Acknowledgments}
P.L.I.~thanks the Department of Physics and Astronomy at Vanderbilt University, Nashville, for the kind hospitality while this work was in progress. P.L.I visit was supported by the University of Bologna, Italy, and Particle and Cosmology BO11 Theory Group (G.~Venturi).  This work was supported by U.S. DoE grant no. DE-FG05-85ER40226.


%
\end{document}